\documentclass[12pt,twoside]{article}
\usepackage{fleqn,espcrc1}

\usepackage{graphicx}
\usepackage[figuresright]{rotating}

\def\gtaprx{ \mathrel{  \vcenter{
                        \offinterlineskip \hbox{$>$}
                        \kern 0.3ex \hbox{$\sim$}    } } }
\def\ltaprx{ \mathrel{  \vcenter{
                        \offinterlineskip \hbox{$<$}
                        \kern 0.3ex \hbox{$\sim$}    } } }

\newcommand{\AmS}{{\protect\the\textfont2
  A\kern-.1667em\lower.5ex\hbox{M}\kern-.125emS}}

\title{Halo Star Abundances and r-Process Synthesis} 

\author{James W. Truran\address[University of Chicago]{Department of Astronomy and 
      Astrophysics, Enrico Fermi Institute, 
      University of Chicago, 5640 S. Ellis Ave., 
      Chicago, IL 60637.},  
      John J. Cowan\address[University of Oklahoma]{Department of Physics and 
      Astronomy, University of Oklahoma, 440 W. Brooks St., Norman, OK 73019.}, 
      and
      Brian D. Fields\address[University of Illinois]{Department of Astronomy,
      University of Illinois, 1011 W. Green St., Urbana, IL 61801.}}
       
\begin{document}

\maketitle

\begin{abstract}
The heavy elements formed by neutron capture processes have an interesting
history from which we can extract useful clues to and constraints
upon the star formation and nucleosynthesis history of Galactic matter.
Of particular interest are the heavy element compositions of extremely
metal-deficient stars.
At metallicities [Fe/H] $\leq$ -2.5, stellar abundance data (for both halo 
field stars and globular cluster stars) provides strong confirmation of the 
occurrence of a robust r-process mechanism for the production of the main 
r-process component, at mass numbers A $\gtaprx$ 130-140. 
The identification of an environment provided by massive stars and
associated Type II supernovae as an r-process site seems compelling.
Scatter in the ratio [r-process/Fe] provides a measure of the level of 
inhomogeneity characteristic of the halo gas at that early epoch. 
Increasing levels of s-process enrichment with increasing
metallicity reflect the delayed contributions from the intermediate mass
stars that provide the site for s-process nucleosynthesis during the AGB phase
of their evolution. For the mass region A $\ltaprx$ 130, the abundance 
patterns in even the most metal deficient stars are not consistent with the 
solar system r-process abundance distribution, providing evidence for the 
fact that the r-process isotopes identified in solar system matter are in 
fact the products of two distinct r-processes nucleosynthesis events. 
 We review recent observational studies of
heavy element abundances in low metallicity stars and explore some
implications of these results for nucleosynthesis and early Galactic chemical
evolution.


\end{abstract}

\section{INTRODUCTION}

Element abundance patterns in very metal-poor halo field stars and globular
cluster stars play a crucial role in guiding and constraining theoretical
models of nucleosynthesis$~$\cite{whe89,mcw95}.
Nowhere is this more true than for the case of the neutron-capture processes
that are understood to be responsible for the synthesis of the bulk of the
heavy elements in the mass region A $\gtaprx$ 60: the s-process and the
r-process. Nucleosynthesis theory identifies quite different astrophysical
sites for these two distinct processes. r-Process nuclei (at least the 
heavy (A $\gtaprx$ 140) r-process isotopes) are
{\it primary} nucleosynthesis products, presumed to have 
been formed in an environment associated
with the evolution of massive stars (M $\gtaprx$ 10 M$_\odot$) to supernova
explosions of Type II and the formation of neutron star remnants.
s-Process nuclei are understood to be products of neutron captures
on preexisting silicon-iron ``seed'' nuclei, occurring both in the helium
burning cores of massive stars and particularly in the thermally pulsing
helium shells of asymptotic giant branch (AGB) stars. In this picture, the
first heavy (A $\gtaprx$ 60) elements introduced into the interstellar gas
component of our Galaxy are expected to have been r-process nuclei formed
in association with massive stars, on timescales $\tau_*$ $<$ 10$^8$ years
$~$\cite{tru81}.

That the general features of this simple model are correct is confirmed by
the finding that r-process contributions dominate the heavy element abundances
in extremely metal-poor halo stars. This issue is discussed in the next
section. The implications of trends in the heavy element abundances as a
function of [Fe/H] for the early star formation history of
the Galaxy and Galactic chemical evolution are then examined.
The interpretation of the observed trends in [Ba/Eu] in the context of 
models of Galactic chemical evolution is discussed in section 3. 
The implications of the increasingly pronounced scatter in the ratio of 
the abundance of r-process nuclei to iron [r-process/Fe] with decreasing 
metallicity are explored in section 4. Evidence of and a possible model for 
a second (weak) r-process nucleosynthesis event$~$\cite{truran00} 
are discussed in section 5. Discussion and conclusions follow. 

\begin{figure}
\centering
\includegraphics[width=.7\textwidth]{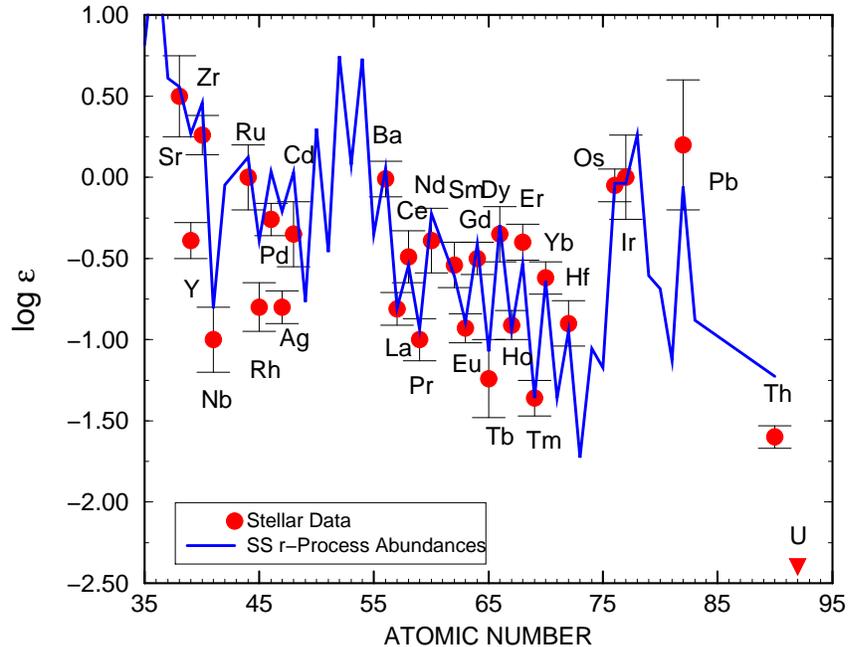}
\noindent
\caption{The heavy element abundance patterns for
CS 22892-052 compared with the solar system
r-process abundance distribution (solid line)$~$\cite{sne00}}
\label{labelfigure1}
\end{figure}

\section{R-PROCESS ABUNDANCES IN METAL-POOR HALO STARS} 

Significant progress in our understanding of the site of operation of
r-process nucleosynthesis has followed from studies of the abundances of
elements from barium through bismuth in metal-poor stars.
Both ground-based$~$\cite{sne96,mcw98,sne00,bur00} and 
space based$~$\cite{cow96,sne98} observational studies have confirmed that 
the abundances of the heavy neutron-capture products (A $\gtaprx$ 130-140) 
in the most metal deficient halo field stars and globular cluster stars 
([Fe/H] $\ltaprx$ -2.5) were formed in a robust r-process event. This is 
most clearly reflected in the abundance pattern for the ultra metal poor 
([Fe/H] = -3.1), but r-process enriched ([r-process/Fe] $\approx$ 30-50), 
halo star CS 22892-052$~$\cite{sne96,sne00} shown in Figure 1. Note the 
truly remarkable agreement of the elemental abundance pattern in the region 
from barium through bismuth with the solar system elemental pattern. 

\begin{figure}
\centering
\noindent
\includegraphics[width=.7\textwidth]{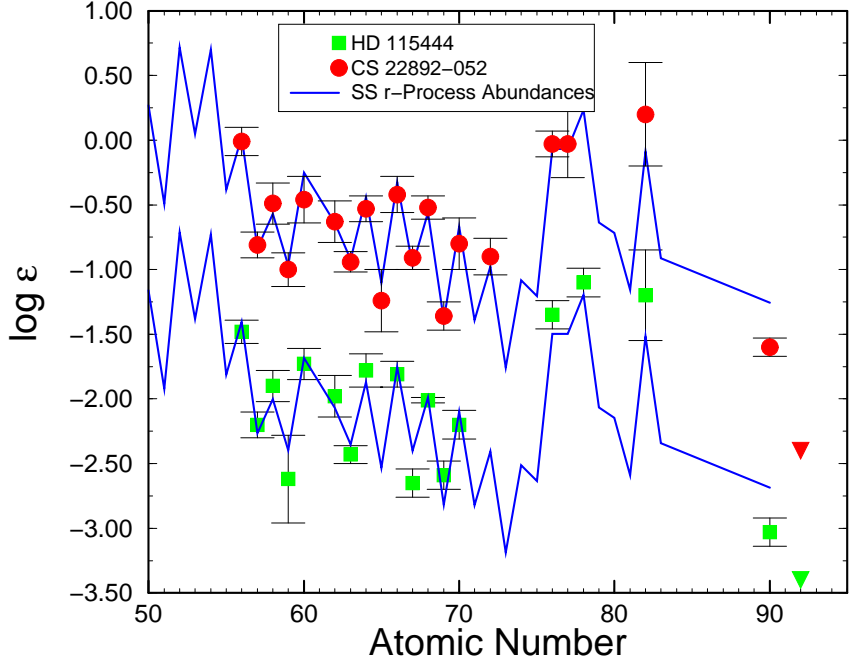}
\caption{The heavy element abundance patterns for the two stars
CS 22892-052 and HD 115444 are compared with the solar system
r-process abundance distribution (solid lines)$~$\cite{sne00,weston00}}
\label{labelfigure2.ps}
\end{figure}

It is also important to recognize that this detailed level of agreement of the 
metal-poor star heavy element abundance pattern with (solar system) r-process 
abundances is a feature of all studied stars of [Fe/H] $\ltaprx$ -2.5. The 
heavy element abundance patterns for the two stars 
CS 22892-052$~$\cite{sne96,sne00} and HD 115444$~$\cite{weston00} are compared 
with the solar system r-process abundances in Figure 2. 

The robustness of the
r-process mechanism operating in the early Galaxy is reflected in the
close agreement of these two stellar patterns (which represent the
nucleosynthesis products of, at most, a relatively small number of
earlier stars) with the solar
system r-process pattern (which represents the accumulated production
of r-process elements over billions of years of Galactic evolution).
Note the extraordinary agreement with solar system r-process abundances
over this range, for the eighteen elements for which abundance data is
available. These data provide conclusive evidence for the operation of
an r-process at the earliest Galactic epochs that synthesizes the heavy
r-process nuclei (barium and beyond), including the long lived isotope
$^{232}$Th critical to dating. The identification with massive stars seems
compelling, although it is possible that neutron star mergers rather than
a supernova environment may be responsible.

\section{CHEMICAL EVOLUTION OF R-PROCESS AND S-PROCESS ABUNDANCES}

Studies of the time history of the heavy (neutron capture) elements
have historically been driven by observations. Early trends in the
s-process element patterns as a function of [Fe/H], 
noted by Pagel$~$\cite{pag67}, 
motivated theoretical consideration of such trends in the context of
models of Galactic chemical evolution$~$\cite{tru71}.
The great increase in abundance data for metal-poor stars over the past
decade has similarly motivated considerable theoretical activity
$~$\cite{mat92,ish99,tra99}.
The ratio [Ba/Eu], which reflects the ratio of s-process to
r-process elemental abundances, is displayed in Figure 3 as a function of
[Fe/H], for a large sample of halo and disk stars from the survey by 
Burris {\it et al.}$~$\cite{bur00}. Note that
at the lowest metallicities the [Ba/Eu] ratio is consistent
with that for pure r-process matter. The subsequent increase in the [Ba/Eu]
ratio with increasing [Fe/H] (i.e. increasing time)
to its solar value
reflects the input to the interstellar medium of the products of
s-process nucleosynthesis in intermediate mass stars of lifetimes
greater than those of the massive star sites of r-process nucleosynthesis.
This data provides important clues to the history of s-process and r-process 
nucleosynthesis over the course of Galactic evolution. 

\begin{figure}
\centering
\noindent
\includegraphics[width=.7\textwidth]{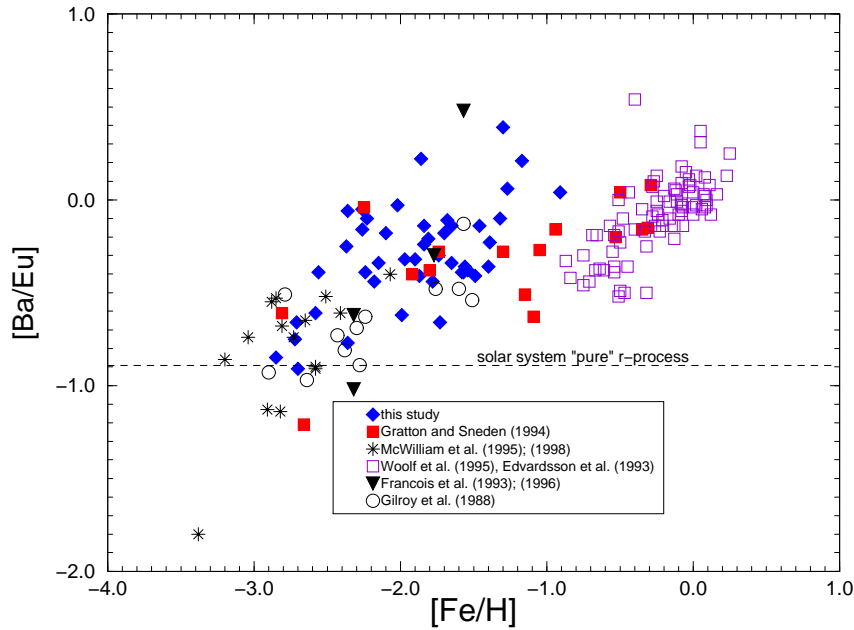}
\caption{The ratio [Ba/Eu], which reflects the ratio of s-process to
r-process elemental abundances, is displayed as a function of
[Fe/H] for a large sample of halo and disk stars$~$\cite{bur00} }
\label{labelfigure3.ps }
\end{figure}

\section{IMPLICATIONS OF SCATTER IN [r-Process/Fe]}

r-Process abundances in low metal stars also provide an important
further clue to the nature of the r-process site.
The ratio [r-process/Fe] is displayed as a function of [Fe/H] for a large
sample of halo and disk stars$~$\cite{bur00} in Figure 4.
The increasing level of scatter of this ratio with decreasing metallicity,
most pronounced at values below [Fe/H] $\sim$ -2.0, makes clear that not
all early stars are sites for the formation of both r-process nuclei and
iron. In the absence of other sources for either iron peak nuclei or
r-process nuclei in the early Galaxy, this scatter is consistent with the
view that the r-process operates in only a small fraction
($\sim$ 1/10th to 1/50th) of the massive stars that produce iron
$~$\cite{fields00}.
Note again that this refers {\it only} to the nucleosynthesis sites for
the heavy r-process isotopes (barium and beyond), and not to any possible
site for the production of the lighter r-process nuclei.

\begin{figure}
\centering
\noindent
\includegraphics[width=.7\textwidth]{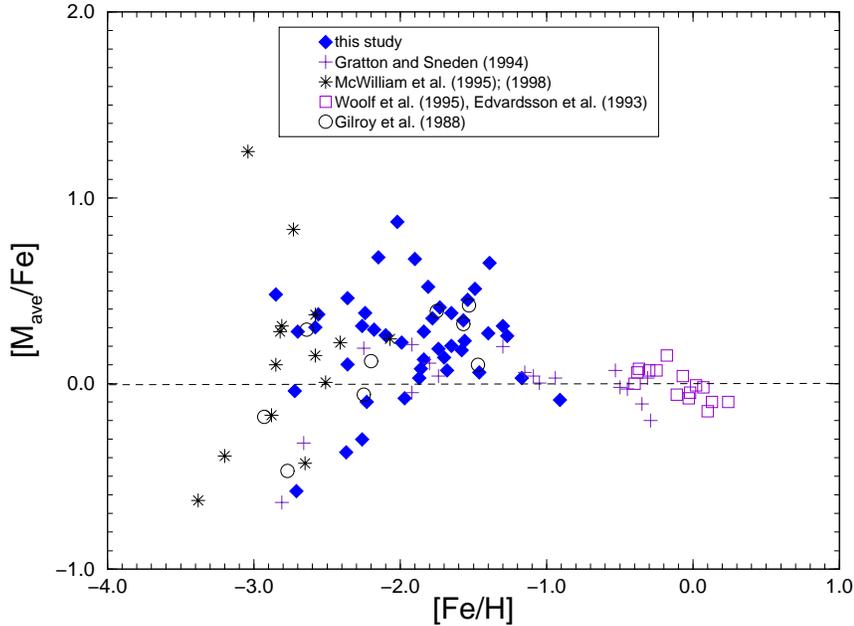}
\caption{The ratio [r-process/Fe] is displayed as a function of
[Fe/H] for a large sample of halo and disk stars$~$\cite{bur00} }
\label{labelfigure4.ps}
\end{figure}

The {\rm scatter} in the $r$/Fe trend is itself interesting
and can teach us much about the formation of the first
supernovae and their spatially {\em inhomogeneous}
enrichment of the halo stars.
The scatter has been studies in detailed
models, by Ishimaru \& Wanajo$~$\cite{ishimaru99}, 
Raiteri et al.$~$\cite{raiteri99}, and 
Argast et al.$~$\cite{argast00}.
The essence of what is going on
is captured in a simple, stripped-down model we have constructed
for the enrichment of the halo stars.

The $r$-process abundance data in extremely metal poor (i.e., low
[Fe/H]) stars drives the model assumptions in a fairly direct
manner.
Namely, (1) from low metallicity of [Fe/H] $\sim -3$ or less, we conclude 
that each star samples only a few (perhaps only one or two?) nucleosynthesis
events 
(Audouze \& Silk$~$\cite{audouze95}; see also Wasserburg, these proceedings).  
(2) The low metallicities also imply short-lived progenitors of both
the Fe and $r$-process observed;
we will refer to these progenitors as supernovae, though
NS-NS binaries are also possible$~$\cite{rosswog00}. 
(3) From the large scatter in observed $r$/Fe, we infer that
not all supernovae make both the (heavy) $r$-process and iron.
(4) Finally, from the stunningly solar
ratios {\em among} the different $r$-process elements
within single stars (see, e.g., our Figure 2), we conclude that
the $r$-process is {\em universal}--i.e., 
the events that produce the $r$-process always closely reproduce the solar 
system r-process pattern. 

We therefore posit that there are two populations of supernovae,
one of which yields mostly the heavy (A $\gtaprx$ 140) 
$r$-process isotopes (more specifically,
a high $r$/Fe ratio), while the other yields iron
(i.e., a low $r$/Fe).  The $r$/Fe yield of each population is
a parameter, but is constrained by the data which sets a lower limit
on $(r/{\rm Fe})_{\rm high}$, and an upper limit on $(r/{\rm Fe})_{\rm low}$.
Furthermore, an additional and key constraint comes from the fact
that, averaged over many events of both types, the final
$r$/Fe ratio has to be that which is observed at [Fe/H] = -1 
(the highest metallicity at which Fe is not contaminated by
Type Ia supernova debris).
This ratio, $(r/{\rm Fe})_{\rm avg} \simeq 3 (r/{\rm Fe})_{\odot}$
sets the fraction $f$ of high $r$/Fe events, since
$r/{\rm Fe}_{\rm avg} = f (r/{\rm Fe})_{\rm high} + (1-f)(r/{\rm Fe})_{\rm low}$.
Also, for simplicity we assume that each supernovae of either population
produces the same iron yield.  

Our model then consists of a Monte-Carlo simulation of
halo-star enrichment by supernovae.
For each star, we impose a total number $N$ of supernova
ancestors; this fixes the [Fe/H] for the star.  
Of these, the numbers of high and low $r$/Fe
progenitors are taken to be random variables.
This is meant to encode the stochastic and inhomogeneous
nature of the halo.  Thus, a star's supernova antecedents
are chosen from a binomial distribution,
whose means are the expected values $fN$ and $(1-f)N$;
this fixes $r$/Fe for the star.  The Poisson noise
in the high/low ancestry leads to scatter in 
$r$/Fe, which scales as $1/\sqrt{N} \propto 1/\sqrt{{\rm Fe/H}}$ and thus increases
with decreasing Fe/H, as observed.  
The results of a typical run are shown in Figure 5, 
which also shows the observed scatter.
We see that this very simple model nicely accounts for
the scatter simply by allowing for stochastic, inhomogeneous
mixing of the $r$-process and iron from two populations.

\begin{figure}
\centering
\noindent
\includegraphics[width=.8\textwidth]{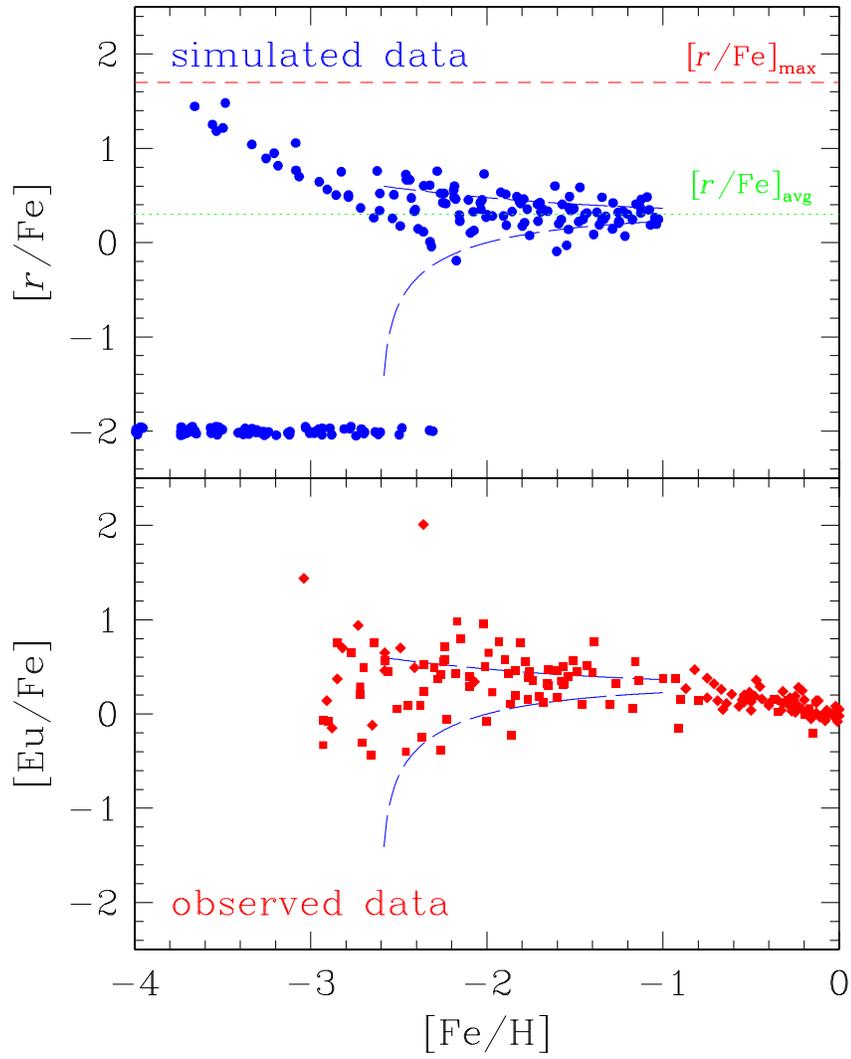}
\caption{
Ratios $r$/Fe as a function of iron abundance.
{\em Upper panel}:  Monte Carlo simulation, upper limits
indicated as points at [$r$/Fe] = -2.  {\em Lower panel}:
observed data.  In both cases, the dashed curves 
indicate the model prediction of the 1-$\sigma$ envelope
in the scatter.} 
\label{labelfigure5.ps}
\end{figure}

\section{EVIDENCE FOR AND A MODEL FOR A SECOND R-PROCESS EVENT}

Recently, Sneden {\it et al.}$~$\cite{sne00} obtained detections of six new
elements in the relatively unexplored elemental regime from Z = 41--48 in
CS 22892--052. These data are displayed in our Figure 1. Note particularly that 
the close agreement of the abundance pattern for  
the heavier n-capture elements with the solar system r-process
pattern does not extend to the lighter elements, {\it e.g. the region below Ba}.
This suggests that, perhaps, a second r-process site may be required. 

It is also interesting to note that there exists evidence for a second,
weak r-process component from another source.
Wasserburg, Busso, \& Gallino$~$\cite{was96}
have pointed out that the abundance levels
of the short-lived isotopes $^{107}$Pd and $^{129}$I in primordial solar
system matter are inconsistent with their having been formed
in uniform production together with the heavy r-process radioactivities
(specifically, $^{182}$Hf, $^{235}$U, $^{238}$U and $^{232}$Th). They suggest
that these two different mass ranges of nuclei require different timescales
for production and therefore suggest two distinct r-process sites.
This seems entirely compatible with the weak (A $\le$ 130-140) and main
(A $\ge$ 130-140) r-process components as we have identified them above.

The appearance of r-process nuclei in the oldest known stars in our Galaxy
strongly suggests the identification of the site of the main r-process
component (A $\gtaprx$ 130-140) with massive stars and associated Type II
supernova environments. We note that the three mechanisms currently
considered for the r-process are all associated in some way with this
environment. This includes nucleosynthesis associated with: (i) neutrino-driven
winds from forming neutron stars$~$\cite{woo94,tak94}; (ii) neutron star
mergers$~$\cite{lat77,fre99}; and (iii) magnetic jets from collapsing stellar
cores$~$\cite{leb70}.

An alternative possible site for r-process synthesis is that associated
with the helium and carbon shells of massive stars undergoing
supernovae$~$\cite{tru78,fkt79,bla81}.
Shock processing of these regions can give rise to
significant neutron production via such reactions as
$^{13}C(\alpha,n)^{16}O$, $^{18}O(\alpha,n)^{21}Ne$, and
$^{22}Ne(\alpha,n)^{25}Mg$, involving residues of hydrostatic burning phases.
The early studies cited above were motivated by the desire to produce the
entire range of r-process nuclei through uranium and thorium. It was found,
however, that this could be accomplished only with the use of excessive and
quite unrealistic concentrations of e.g. $^{13}$C$~$\cite{cow85}.

\begin{figure}
\centering
\includegraphics[width=.8\textwidth]{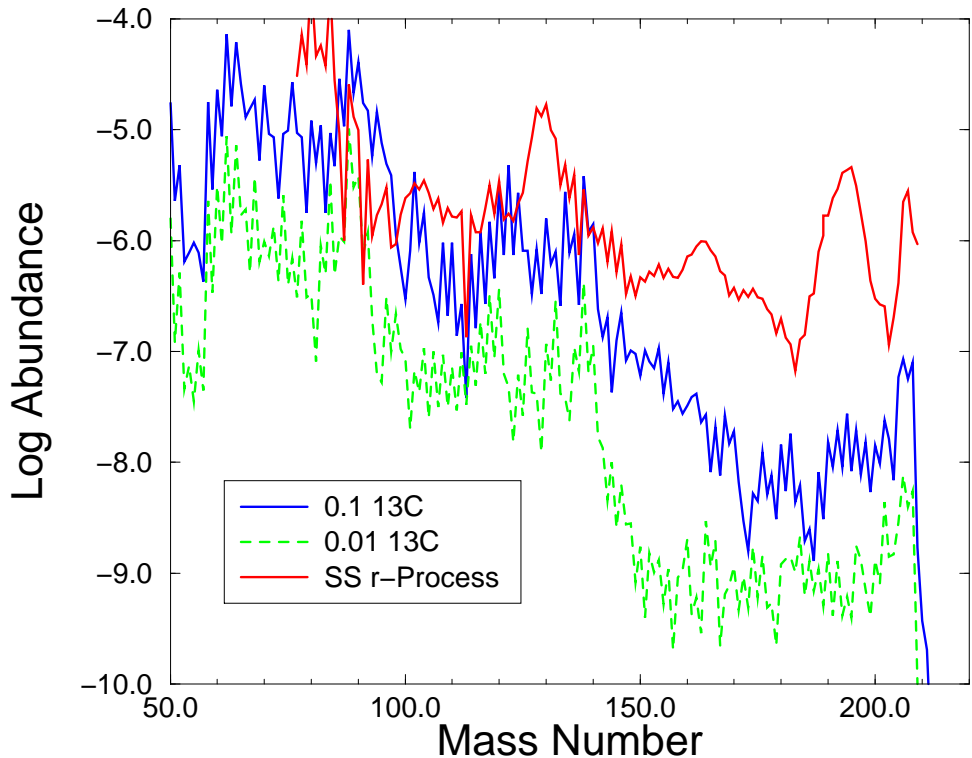}
\caption{The r-process abundance patterns resulting from
two representative $^{13}$C($\alpha$,n)$^{16}$O
neutron exposures in the helium shell
of a massive star are compared with the solar system
r-process abundance distribution (solid line).}
\label{labelfigure6}
\end{figure}

The observational and theoretical considerations discussed previously
serve to ease these demands: we need now only to be able to produce the
r-process nuclei in the mass range through A $\approx$ 130-140. Guided by
stellar models for massive stars at the end of their hydrostatic
evolution$~$\cite{woo95}, we have calculated the r-process history of matter
processed through the shock heated helium shell for two choices of the
initial abundance of $^{13}$C: X$_{13}$=0.1 and 0.01. The initial (post shock)
temperature and density were taken to be 7x10$^8$ $^o$K and 10$^4$ g/cc,
respectively. Here we have assumed an active $^{13}$C($\alpha$,n)$^{16}$O
neutron source acting on a preexisting solar abundance pattern as seed
nuclei, and calculated nucleosynthesis occurring on an expansion time scale.
(We note that we might also expect important contributions from the
$^{18}O(\alpha,n)^{21}Ne$ and $^{22}Ne(\alpha,n)^{25}Mg$ reactions.)
The results are displayed in Figure 6, where the calculated abundance
patterns for the two choices of X$_{13}$ are compared with the solar system
r-process pattern. We note that for the more realistic case of X$_{13}$=0.01,
the element abundance enrichments fall sharply in the range
A $\gtaprx$ 130. The r-process products formed under these conditions 
are thus confined to the light
mass region that seems not to receive significant contributions from the
main r-process component.

\section{DISCUSSION AND CONCLUSIONS}

r-Process abundances in low metal stars provide several important clues to
the nature of the r-process site:

\begin{itemize}

\item{} r-Process synthesis of A $\gtaprx$ 130-140 isotopes happens early
in Galactic history$~$\cite{tru81},
prior to input from AGB stars to the abundances of
heavy s-process isotopes.

\item{} The chemical evolution of the main sources of s-process and r-process 
nuclei is reflected in the evolution of [Ba/Eu] with respect to [Fe/H], 
shown in Figure 3$~$\cite{bur00}. The 
stellar data indicate that, at metallicities [Fe/H] $\ltaprx$ -2.5, the 
abundance pattern of the heavy {\em n}-capture products is consistent with 
purely {\em r}-process nucleosynthesis production. {\em s}-process 
contributions are first seen in some stars at metallicities as low as 
[Fe/H] $\sim$ -2.75, and are observed in most stars with metallicities 
[Fe/H] $>$ -2.3. This identifies the delayed entry of Ba from the low mass 
AGB stars that represent the site of operation of the main {\em s}-process  
component, on a timescale $\sim$ 10$^9$ years. 

\item{} The observed scatter in the ratio [r-Process/Fe] is quite large at 
low metallicities$~$\cite{bur00}, while at high metallicities it essentially disappears. 
We have shown that this behavior can be understood as a reflection of the 
increasing level of inhomogeneity of Galactic matter at earlier epochs. 

\item{} The r-process mechanism for the synthesis of the A $\gtaprx$ 130-140
isotopes (which we will refer to as the `main' component)
is extremely robust. This is reflected in the fact that the abundance patterns
in the most metal deficient (oldest) stars, which may have received
contributions from only one or a few r-process events, are nevertheless
entirely
consistent with the r-process abundance pattern which characterizes solar
system matter.
(The abundance patterns for the two stars CS 22892-052$~$\cite{sne96,sne00}
and HD 115444$~$\cite{weston00} shown in Figure 2 reveal this  
remarkable agreement for two stars of low metallicity but high [r-Process/Fe].)

\item{} The abundance pattern
in the mass regime below A $\approx$ 130 does not exhibit this
consistency.
It is the source of these lighter r-nuclei
(the `weak' component) that we have sought to identify in our very 
preliminary survey of r-process nucleosynthesis conditions in the helium and 
carbon shells of Type II supernovae$~$\cite{truran00}.  

\item{}
Our numerical results confirm that the helium and 
carbon layers of massive stars, when
subjected to outgoing shocks characteristic of Type II supernovae, can
represent an important nucleosynthesis site - specifically for the light
r-process nuclei. The great sensitivity to conditions of temperature, density,
and initial composition that we have found in our exploratory survey strongly
suggest that this r-process mechanism will prove less robust in its ability
to reproduce the observed r-process pattern in the mass
range A $\ltaprx$ 130-140.

\end{itemize}

This research was funded in part by NSF grants AST-9618332 and AST-9986974 
(JJC), by grant GO-08342 from the Space Science Telescope Institute, 
and by the Department of Energy, under Grant No. B342495 to the Center for 
Astrophysical Flashes at the University of Chicago.

\end{document}